\documentclass{jfm}
\usepackage{graphicx}
\usepackage{epstopdf, epsfig}
\usepackage{amsmath}

\usepackage[colorlinks, allcolors=blue]{hyperref} 

\newcommand\Ren{\mbox{\textit{Re}}}  
\newcommand\Ra{\mbox{\textit{Ra}}}
\newcommand\Cn{\mbox{\textit{Cn}}}

\newcommand\We{\mbox{\textit{We}}}
\newcommand\Fr{\mbox{\textit{Fr}}}

\newcommand\Nun{\mbox{\textit{Nu}}}
\newcommand\Prn{\mbox{\textit{Pr}}} 
\newcommand\Pe{\mbox{\textit{Pe}}}

\title{Heat transfer in turbulent Rayleigh-B\'enard convection within two immiscible fluid layers}

\author{Hao-Ran Liu\aff{1},
  Kai Leong Chong\aff{2,1},\corresp{\email{klchong@shu.edu.cn}},
   Rui Yang\aff{1},
  Roberto Verzicco\aff{3,4,1}
 \and Detlef Lohse\aff{1,5},\corresp{\email{d.lohse@utwente.nl}}}

\affiliation{\aff{1}Physics of Fluids Group and Max Planck Center Twente for Complex Fluid Dynamics,\\
MESA+Institute and J. M. Burgers Centre for Fluid Dynamics, University of Twente,\\
P.O. Box 217, 7500AE Enschede, The Netherlands
\aff{2}Shanghai Key Laboratory of Mechanics in Energy Engineering, Shanghai Institute of Applied Mathematics and Mechanics, School of Mechanics and Engineering Science, Shanghai University, Shanghai, 200072, PR China
\aff{3}Dipartimento di Ingegneria Industriale, University of Rome ``Tor Vergata", Via del Politecnico 1, Roma 00133, Italy
\aff{4}Gran Sasso Science Institute - Viale F. Crispi, 7 67100 L’Aquila, Italy
\aff{5}Max Planck Institute for Dynamics and Self-Organization, Am Fassberg 17, 37077 Göttingen, Germany}

\begin{document}

\maketitle

\begin{abstract}
We numerically investigate turbulent Rayleigh-B\'enard convection within two immiscible fluid layers, aiming to understand how the layer thickness and fluid properties affect the heat transfer (characterized by the Nusselt number $\Nun$) in two-layer systems. Both two- and three-dimensional simulations are performed at fixed global Rayleigh number $\Ra=10^8$, Prandtl number $\Prn=4.38$, and Weber number $\We=5$. We vary the relative thickness of the upper layer between $0.01 \le \alpha \le 0.99$ and the thermal conductivity coefficient ratio of the two liquids between $0.1 \le \lambda_k \le 10$. Two flow regimes are observed: In the first regime at $0.04\le\alpha\le0.96$, convective flows appear in both layers and $\Nun$ is not sensitive to $\alpha$. In the second regime at $\alpha\le0.02$ or $\alpha\ge0.98$, convective flow only exists in the thicker layer, while the thinner one is dominated by pure conduction. In this regime, $\Nun$ is sensitive to $\alpha$. To predict $\Nun$ in the system in which the two layers are separated by a unique interface, we apply the Grossmann-Lohse theory for both individual layers and impose heat flux conservation at the interface. Without introducing any free parameter, the predictions for $\Nun$ and for the temperature at the interface well agree with our numerical results and previous experimental data. 

\end{abstract}


\section{Introduction}

Turbulent flows with two layers are ubiquitous in nature and technology: for example, the coupled flows between the ocean and the atmosphere \citep{neelin1994arfm}, the convection of the Earth's upper and lower mantles \citep{tackley2000science}, oil slicks on the sea \citep{fay1969book} and even cooking soup in daily life. Compared to flow with only a single layer, multilayer flow has more complicated flow dynamics and thus more complicated global transport properties. These are determined by the viscous and thermal coupling between the fluid layers. In order to investigate the heat transport in two-layer thermal turbulence, we choose turbulent Rayleigh-B\'enard convection as canonical model system. The main question to answer is: How does the global heat transport depend on the ratios of the fluid properties and the thicknesses of the two layers?

Rayleigh-B\'enard (RB) convection, where a fluid is heated from below and cooled from above, is the most paradigmatic example to study global heat transport in turbulent flow (see the reviews of \cite{ahl09,loh10,chi12}). Also previous studies on multiphase flow have focused on RB convection as model system, for example, the studies on boiling convection \citep{zho09,lak13,wang2019nc-ziqi}, ice melting in a turbulent environment \citep{wang2021pnas}, or convection laden with droplets \citep{liu2021prl,pelusi2021sm}. These are mainly dispersed multiphase flow. Here we focus on two-layer RB convection, which is highly relevant in geophysics and industrial applications.

Most studies on two-layer RB convection have hitherto been conducted in the non-turbulent regime \citep{nataf1988jp,prakash1994ijmf,busse2009pre,diwakar2014jfm}, while only a few studies focused on the turbulent one: \cite{davaille1999nature} experimentally studied RB convection with two miscible layers and displayed the temperature profiles in each layer and the deformation of the interface. Experiments of the RB convection with two immiscible fluid layers were reported by \cite{xie2013jfm}, who quantified the viscous and thermal coupling between the layers and showed that the global heat flux weakly depends on the type of coupling, i.e., the heat flux at viscous coupling normalized by that at thermal coupling varies from $0.997$ to $1.004$.

Next to the experiments, the numerical study of \cite{yoshida2016pof} focused on the heat transport efficiency in two-layer mantle convection with large viscosity contrasts. Our prior numerical simulations \citep{liu2021jfm} explored a wide range of Weber numbers and density ratios between the two fluids and revealed the interface breakup criteria in two-layer RB convection. Also most previous studies have focused on the flow dynamics and behavior of the interface between the two layers. The issue of global heat transport is still quite unexplored, especially in the turbulent regime.

In the present work, we numerically investigate RB convection within two immiscible fluid layers by two- and three-dimensional direct numerical simulations. We calculate the global heat transport for various layer thicknesses and fluid properties. In addition, we extend the Grossmann-Lohse (GL) theory for standard RB convection to the two-layer system and are able to predict the heat transport of the system. The GL theory \citep{gro00,gro01,gro02,ste13} is a unifying theory for the effective scaling in thermal convection, which successfully describes how the Nusselt number $\Nun$ and the Reynolds number $\Ren$ depend on the Rayleigh number $\Ra$ and the Prandtl number $\Prn$. This theory has been well validated over a wide parameter range by many single-phase experimental and numerical results \citep{ste13}. Here, we further show that the predictions from the GL theory can be extended to two-layer systems without introducing any new free parameter, and the extended theory well agrees with our numerical results and with previous experimental data \citep{wang2019nc-ziqi}.

The organization of this paper is as follows: The numerical method and setup are introduced in Section \ref{meth}. The main results are presented in Sections \ref{sec1}$-$\ref{sec4}, namely the flow structure and heat transfer in the two-layer RB convection in Section \ref{sec1}, the temperature at the interface in Section \ref{sec3}, and the effects of the interface deformability on the heat transfer in Section \ref{sec4}. The paper ends with conclusions and an outlook.

\section{Numerical method and setup}\label{meth}

Two- and three-dimensional (2D and 3D) simulations are performed in this study. We consider two immiscible fluid layers (fluid 2 over fluid 1) placed in a domain of dimensions $2H\times2H\times H$ in 3D and $2H\times H$ in 2D with $H$ being the domain height. The numerical method \citep{liu2021jcp} used here combines the phase-field method \citep{jaqcmin1999jcp, ding2007jcp, liu2015jcp} and a direct numerical simulation solver for the Navier-Stokes equations \citep{ver96, poe15cf}, namely \href{https://github.com/PhysicsofFluids/AFiD}{AFiD}. This method has been well validated by our previous study \citep{liu2021jcp}.

In the phase field method, which is widely used in multiphase simulations of laminar \citep{li2020jcp,chen2020jfm} and turbulent flows \citep{roccon2019jfm,roccon2021jfm}, the interface is represented by contours of the volume fraction $C$ of fluid 1. The corresponding volume fraction of fluid 2 is $1-C$. The evolution of $C$ is governed by the Cahn-Hilliard equation,
\begin{equation}
\frac {\partial C} {\partial t} + \nabla \cdot ({\bf u} C) = \frac{1}{\Pe}\nabla^2 \psi,
\label{ch}
\end{equation} 
where $\bf u$ is the flow velocity, and $\psi= C^{3} - 1.5 C^{2}+ 0.5 C  -\Cn^{2} \nabla^2 C$ the chemical potential. We set the P\'eclet number $\Pe=0.9/\Cn$ and the Cahn number $\Cn=0.75h/H$, where $h$ is the mesh size. We take $\Pe$ and $\Cn$ according to the criteria proposed in \cite{ding2007jcp,yue2010jfm,liu2015jcp}, which approaches the sharp-interface limit and leads to well-converged results with mesh refinement. More numerical details, validation cases and convergence tests can be found in \cite{liu2021jcp}.

The flow is governed by the Navier-Stokes equation, the heat transfer equation, and the incompressibility condition,
\begin{equation}
\tilde{\rho} \left(\frac {\partial {\bf u}} {\partial t} + {\bf u} \cdot \nabla {\bf u}\right)= - \nabla P + \sqrt {\frac{\Prn}{\Ra}} \nabla \cdot [\tilde{\mu} (\nabla {\bf u}+ \nabla {\bf u}^{T})] + {{\bf F}_{st}}+{\bf G},
\label{ns}
\end{equation}

\begin{equation}
\tilde{\rho}\tilde{c_p} \left(\frac{\partial {\theta}}{\partial t} + {\bf u} \cdot \nabla \theta \right) =   \sqrt{\frac{1}{ \Prn \Ra }} \nabla \cdot (\tilde{k} \nabla \theta),
\label{t}
\end{equation}

\begin{equation}
\nabla \cdot {\bf u}= 0,
\label{con}
\end{equation}
where $\theta$ is the dimensionless temperature, ${\bf F}_{st}=6\sqrt{2}\psi \nabla C / (\Cn \We)$ the dimensionless surface tension force and ${\bf G}=\left\{[C+\lambda_\beta \lambda_\rho (1-C)] \, \theta-\tilde{\rho}/\Fr\right\} {\bf z}$ the dimensionless gravity. All dimensionless fluid properties (indicated by $\tilde{q}$) are defined in a uniform way,
\begin{equation}
\tilde{q}=C+\lambda_q(1-C),
\end{equation}
where $\lambda_q=q_2/q_1$ is the ratio of the material properties of fluid 2 and fluid 1, marked by the subscripts $2$ and $1$, respectively. The global dimensionless parameters controlling the flow are the Rayleigh number $Ra=\beta_1 g H^3 \Delta/(\nu_1 \kappa_1)$, the Prandtl number $Pr=\nu_1/\kappa_1$, the Weber number $\We=\rho_1 U^2 H/\sigma$, and the Froude number $\Fr=U^2/(gH)$, where $\beta$ is the thermal expansion coefficient, $g$ the gravitational acceleration, $\Delta$ the temperature difference between the bottom and top plates, $\nu=\mu/\rho$ the kinematic viscosity, $\rho$ the density, $\mu$ the dynamic viscosity, $\kappa=k/(\rho c_p)$ the thermal diffusivity, $k$ the thermal conductivity, $c_p$ the specific heat capacity, $U=\sqrt{\beta_1 g H \Delta}$ the free-fall velocity, and $\sigma$ the surface tension of the interface between the two fluids. The heat transfer of the system is characterized by the Nusselt number $\Nun=Q/(k_1\Delta/H)$, with $Q$ being the heat flux.

In this study, we investigate how the heat transfer is influenced by the relative layer thickness $\alpha$ (for the upper layer) and the thermal conductivity coefficient ratio $\lambda_k$ of the upper layer over the lower layer. We vary $\alpha$ from $0$ (pure fluid 1) to $1$ (pure fluid 2), and $0.1\le\lambda_k\le10$, which is in an experimentally accessible range, e.g., $\lambda_k\approx10$ in the experiments by \cite{xie2013jfm} and $\lambda_k\approx8$ in \cite{wang2019nc-ziqi}. For numerical simplicity, we fix the global parameters: $\Ra=10^8$, at which there is already considerable turbulent intensity, $\Prn=4.38$, based on the property of water, and $\We=5$, a value ensuring that the surface tension is still strong enough so that the interface does not to break up. The second term $-\tilde{\rho}/\Fr$ of the gravity $\bf G$ in eq. (\ref{ns}) only determines which fluid is in the lower layer. It has no effect on the flow in each individual layer due to the Boussinesq approximation. We take $\Fr=1$ for numerical convenience. The density ratio $\lambda_\rho=0.8$, reflecting that fluid 2 in the upper layer is less dense than fluid 1 in the lower layer. The other properties are the same in both fluids, for simplicity.

We set the boundary conditions on the plates as $\partial C/\partial z=0$, 
no-slip velocities, and fixed temperature $\theta=0$ (top) and $1$ (bottom), and use periodic conditions in the horizontal directions.  Uniform grids with $1152 \times 1152 \times 588$ gridpoints in 3D cases and $1152 \times 588$ gridpoints in 2D cases are used for the velocity and temperature field, and $4608\times4608\times2304$ gridpoints in 3D cases and $4608\times2304$ gridpoints in 2D cases for the phase field. 
The mesh is sufficiently fine and compares to corresponding single-phase studies in 2D and 3D \citep{ste10,poe13,zhou2017jfm}. 

\begin{figure}
\centering
\includegraphics[width=0.9\linewidth]{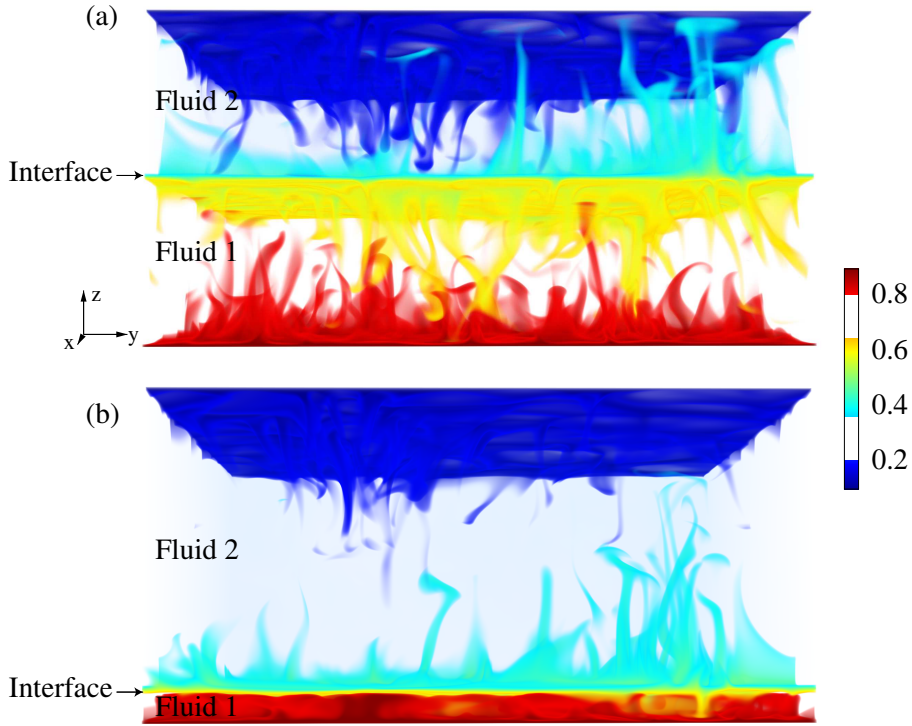}
\caption{\label{3d} Volume rendering of a snapshot of the thermal structures in two-layer Rayleigh-B\'enard convection in three-dimensional simulations at $\Ra=10^8$, $\We=5$, the thermal conductivity coefficient ratio $\lambda_k=1$, and the thickness of the upper layer (a) $\alpha=0.5$ and (b) $\alpha=0.9$. The temperature field is color coded.}
\end{figure}

\section{Flow structure and heat transfer in the two-layer system}
\label{sec1}
We first examine the typical flow structure in turbulent Rayleigh-B\'enard convection with two immiscible fluid layers. Since the Weber number is sufficiently small ($\We=5$), strong enough surface tension prevents the fluids interface from producing droplets, and a clear single interface can be distinguished. 

\begin{figure}
\centering
\includegraphics[width=0.98\linewidth]{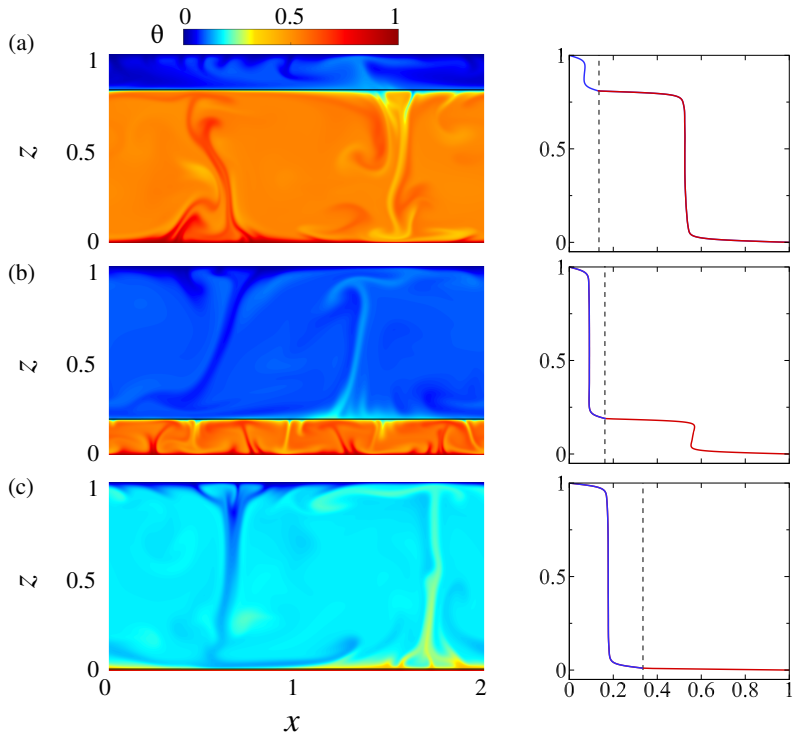}
\caption{\label{2d} Snapshots of the two-dimensional simulations at $\lambda_k=10$ and (a) $\alpha=0.19$, (b) $\alpha=0.81$ and (c) $\alpha=0.99$. The corresponding average temperature profiles are shown as blue (upper layer) and red (bottom layer) curves on the right column, where the dashed lines denote the temperature at the interface. 
}
\end{figure}

As shown in figure \ref{3d} for 3D cases and figure \ref{2d}(a) and (b) for 2D cases, a large-scale circulation exists in each individual layer, which is consistent with previous two-layer studies \citep{davaille1999nature, xie2013jfm, liu2021jfm}. From figure \ref{2d}, we also observe that the temperature profile in each layer resembles that in the single-phase flow, namely the temperature remains constant in the bulk of each layer and strongly varies in the boundary layers near the plates and the interface. 

When one layer becomes thin enough, we observe a temperature inversion therein (see figure \ref{2d}b). A similar feature has been observed in \cite{blass2021}, who considered sheared thermal convection, and in many prior papers \citep{bel93,til93}. The reason for the inversion is that the fluctuations in the flow are not strong enough to fully mix the bulk. Then the remaining hot (cold) fluid is carried downward (upward) by the flow. When one layer is sufficiently thin, pure thermal conduction instead of convection is observed. This is the case once the local $\Ra$ (defined with the layer thickness and its material parameters) becomes smaller than the onset value of convection. For example, in figure \ref{2d}(c), the bottom layer has the local Rayleigh number $\Ra_1=34.4$ and a pure conductive layer is found. Thus, in summary, two regimes are identified: The regime with both convective layers appears at $0.04\le\alpha\le0.96$, and the regime with one conductive layer and one convective layer at $\alpha\le0.02$ or $\alpha\ge0.98$.

\begin{figure}
\centering
\includegraphics[width=1\linewidth]{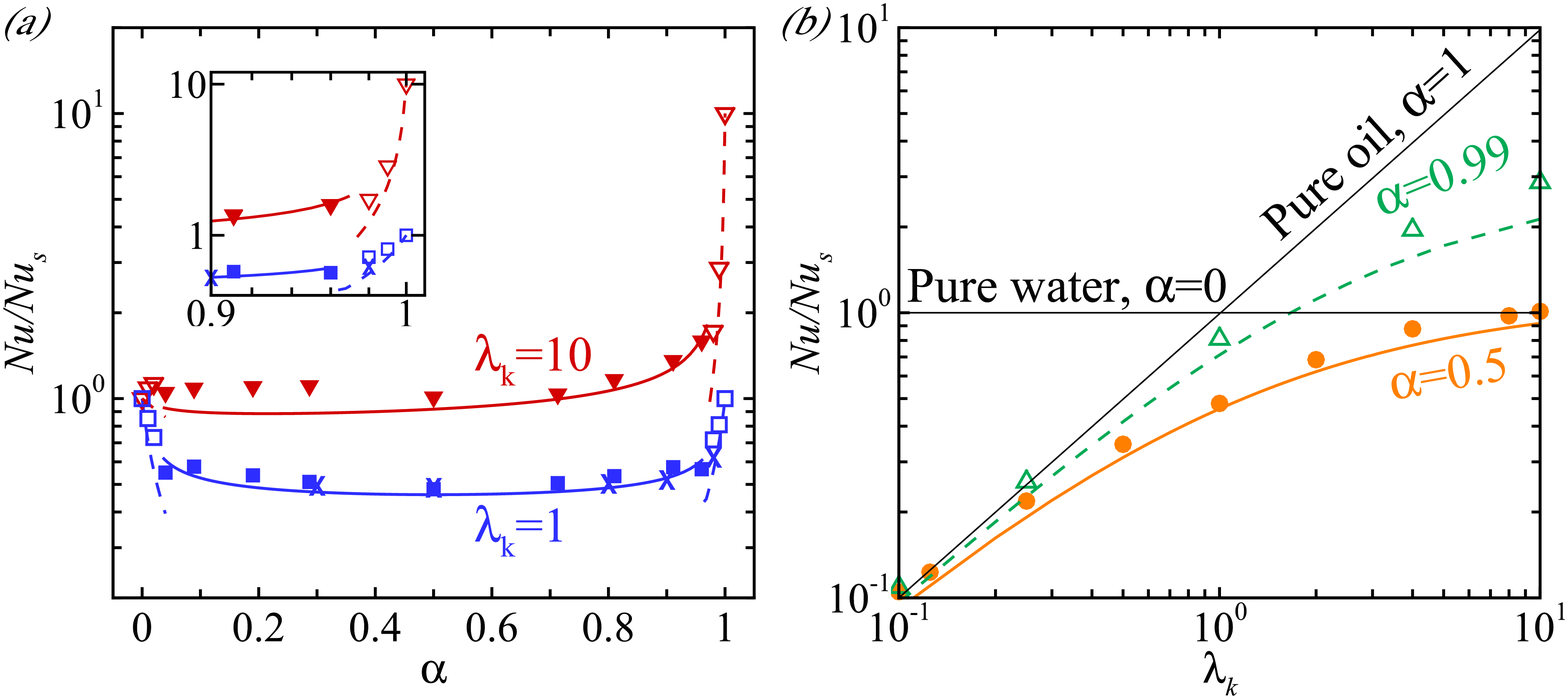}
\caption{\label{nu}   Global Nusselt number $\Nun$ normalized by $\Nun_s$ of the single-phase system as function of (a) $\alpha$ and (b) $\lambda_k$ at $\Ra=10^8$ and $\We=5$. The symbols denote the numerical results and the lines the Grossmann-Lohse theory for two-layer system (\ref{gl1})$-$(\ref{gl5}) at fixed $\lambda_k=1$ and $10$ in (a) and $\alpha=0.5$ and $0.99$ in (b), where the filled symbols/solid lines denote the regime with both convective layers, and the empty symbols/dashed lines the regime with one convective layer and one conductive layer. The three-dimensional cases are marked as cross symbols. The inset in (a) shows a zoom-in view for $\alpha$ close to $1$.}
\end{figure}

In the two distinct regimes, the heat transfer shows different $\alpha$ dependence as shown in figure \ref{nu}(a), where the heat flux is normalized with the heat flux $\Nun_s$ for the single-phase flow (pure fluid 1). In the regime with both convective layers, the heat transfer only minimally varies with $\alpha$ although the layer thickness $\alpha$ changes considerably. More specifically, we obtain $0.48\le\Nun/\Nun_s\le0.57$ over $0.04\le\alpha\le0.96$ for $\lambda_k=1$ while $1.00\le\Nun/\Nun_s\le1.59$ for $\lambda_k=10$. In contrast, in the regime with one conductive layer, $\Nun/\Nun_s$ increases largely from $0.72$ to $1.00$ for $\lambda_k=1$ and from $1.72$ to $10.00$ for $\lambda_k=10$ with $\alpha$ only increasing from $0.98$ to $1$. Note that this interesting dependence of $\Nun/\Nun_s$ on $\alpha$ also holds for the 3D cases.

Also the $\lambda_k$ dependence is examined and shown in figure \ref{nu}(b). As expected, the heat transfer in the two-layer system is limited by the layer with the lower thermal conductivity coefficient.

Obviously, in two-layer RB convection, $\Nun$ is affected by more parameters (namely $\alpha$ and the various ratios between the material properties such as $\lambda_k$) as compared to the single-phase case. To predict $\Nun$ in the two-layer system, we regard the convection in each layer as single-phase convection. Then we apply the GL theory \citep{gro00,gro01,gro02,ste13} in each individual layer, 
\begin{align}
(\Nun_j-1)\Ra_j\Prn_j^{-2}=&c_1\frac{\Ren_j^2}{g(\sqrt{\Ren_c/\Ren_j})}+c_2\Ren_j^3,\label{gl1}\\
\Nun_j-1=&c_3\Ren_j^{1/2}\Prn_j^{1/2}\left\{f\left[\frac{2a\Nun_j}{\sqrt{\Ren_c}}g\left(\sqrt{\frac{\Ren_c}{\Ren_j}}\right)\right]\right\}^{1/2}+ \notag\\
&c_4\Prn_j\Ren_j f\left[\frac{2a\Nun_j}{\sqrt{\Ren_c}}g\left(\sqrt{\frac{\Ren_c}{\Ren_j}}\right)\right],
\label{gl2}
\end{align}
for $j=1$ and $2$, indicating the fluid $1$ and $2$. Here the local Rayleigh numbers in the respective layers are $\Ra_1=(1-\alpha)^3 (1-\theta_i) \Ra$ and $\Ra_2=\alpha^3 \theta_i \Ra$ with $\theta_i$ being the temperature at the interface, and the local Prandtl numbers are $\Prn_1=\nu_1/\kappa_1$ and $\Prn_2=\nu_2/\kappa_2$. $f$ and $g$ are crossover functions, which are given by \cite{gro00,gro01}. We take the values of the parameters of the theory as given by \cite{ste13}, namely $c_1=8.05$, $c_2=1.38$, $c_3=0.487$, $c_4=0.0252$, and $\Ren_c=(2c_0)^2$ with $c_0=0.922$. These are based on various numerical and experimental data over a large range of $Ra$ and $Pr$ numbers. 
The above equations can be closed by imposing the heat flux balance at the interface,
\begin{equation}
Q_2=Q_1.
\label{gl3}
\end{equation}
These heat fluxes are differently non-dimensionalized as Nusselt numbers, namely
\begin{align}
\Nun_2&=\frac{Q_2}{k_2\frac{\theta_i \Delta}{\alpha H}},\quad\notag\\
\Nun_1&=\frac{Q_1}{k_1\frac{(1-\theta_i) \Delta}{(1-\alpha)H}}.\label{gl5}
\end{align}
With eqs. (\ref{gl1})$-$(\ref{gl5}) we obtain the GL-predictions for $\Nun(\Ra,\Prn)$ for the system with two convective layers.

Just as in single layer RB, the regime with one conductive layer has to be treated differently, instead of simply using (\ref{gl1}) and (\ref{gl2}). First, to determine the criteria for the occurrence, we use eqs. (\ref{gl1})$-$(\ref{gl5}) to calculate $\Ra_1$ and $\Ra_2$, and then compare them to the onset value of convection. Note that the conductive layer only occurs when the layer is thin enough, and in that case the aspect ratio of the conductive layer is much larger than $1$, allowing us to take the onset value to be $1708$ \citep{cha81}. Next, if one of the local Rayleigh numbers is smaller than $1708$, we use $\Nun_j=1$ to replace (\ref{gl1}) and (\ref{gl2}). We get the critical layer thickness $\alpha_c=0.965$ for $\lambda_k=1$ ($\alpha_c=0.972$ for $\lambda_k=10$) at $\Ra_1=1708$ and $\alpha_c=0.035$ for $\lambda_k=1$ ($\alpha_c=0.057$ for $\lambda_k=10$) at $Ra_2=1708$.

In figure \ref{nu}, we compare $\Nun/\Nun_s$ from the simulations with the predictions above. In both regimes, the dependence of $\Nun/\Nun_s$ on $\alpha$ and $\lambda_k$ well agrees with the predictions of the GL theory for the two-layer system. Note that no new fitting parameter was introduced at all.

To further validate our predictions, we compare the theoretical prediction also with previous experiments by \cite{wang2019nc-ziqi}, who measured the heat transfer in RB convection with a water layer over a heavy oil (\href{https://www.acota.co.uk/wp-content/uploads/2018/11/3M-Novec-Fluids-for-organic-rankine-cycle-systems.pdf}{HFE-7000}) layer at the parameters $\Ra\approx4.5\times10^{10}$ and $\Prn\approx8.8$ (calculated for pure water), and $\lambda_\beta\approx0.044$, $\lambda_\nu\approx3.8$, $\lambda_\kappa\approx3.3$, and $\lambda_k\approx7.7$ (water over oil). In the system with $99\%$ water and $1\%$ oil in volume, \cite{wang2019nc-ziqi} measured $\Nun\approx102$ (from figure 3e of that paper). 
With the extended GL theory for the two layers (\ref{gl1})$-$(\ref{gl5}), we obtain $\Nun=90.4$ ($\theta_i=0.53$), which reasonably agrees with the experimental results.

\section{Temperature at the interface}
\label{sec3}
\begin{figure}
\centering
\includegraphics[width=0.85\linewidth]{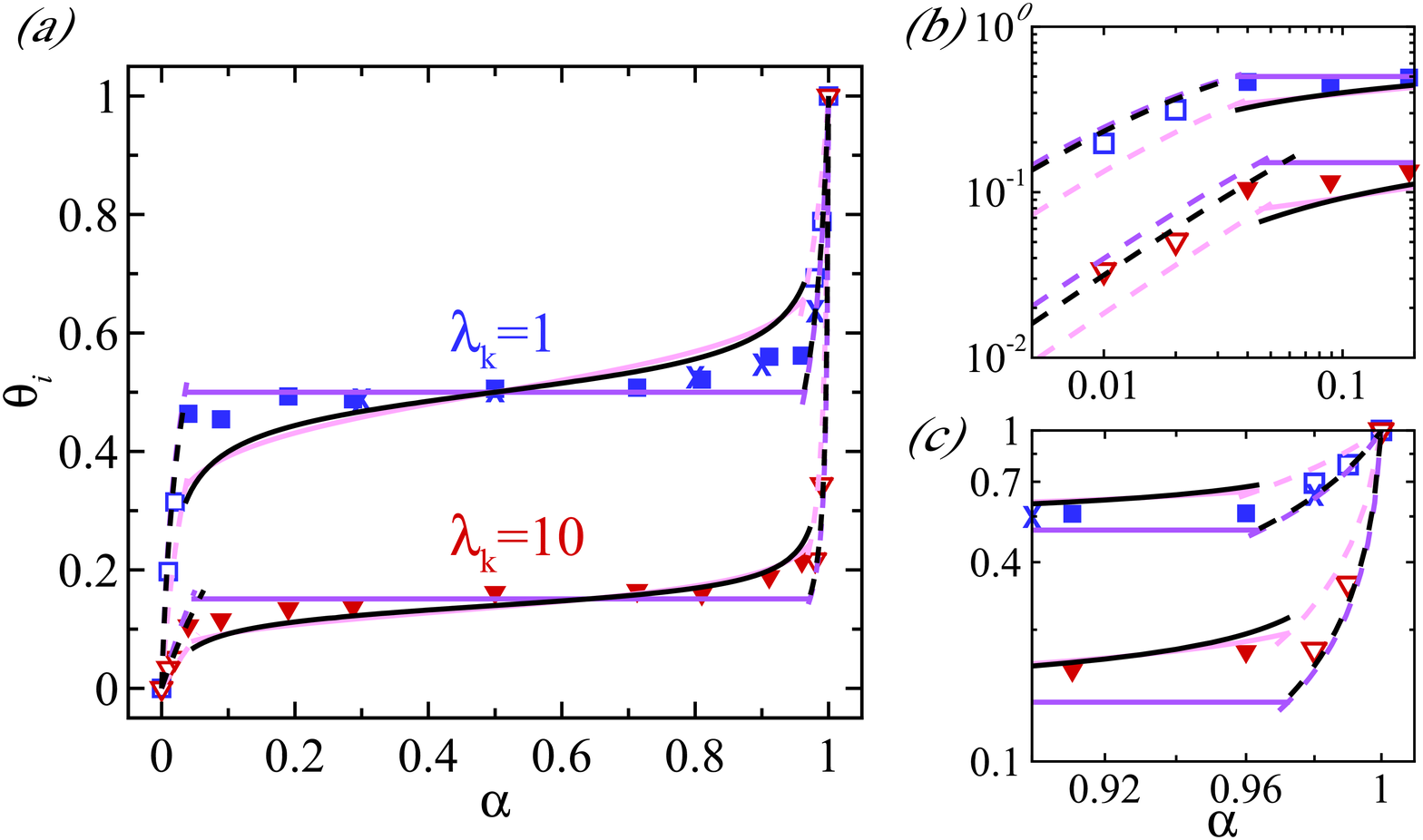}
\caption{\label{ti-a} Temperature at the interface $\theta_i$ as function of $\alpha$. (b) and (c) are zoom-in views of (a). The lines represent the GL theory for the two-layer system (black) in (\ref{gl1})$-$(\ref{gl5}), and the approximate theory in (\ref{scale0})$-$(\ref{scale3}) with $\Nun$ vs $\Ra$ scaling exponent of $\gamma=1/4$ (pink) and $\gamma=1/3$ (purple). The symbols and line types are the same as in figure \ref{nu}. 
}
\end{figure}

The temperature $\theta_i$ at the interface plays a crucial role in the heat transfer of the two-layer system, since it determines the value of the local Rayleigh numbers as mentioned in section \ref{sec1}. We plot the $\alpha$ dependence of $\theta_i$ in figure \ref{ti-a}. Similarly to the $\alpha$ dependence of $\Nun/\Nun_s$, we again observe that $\theta_i$ is insensitive to the change in $\alpha$ over $0.04\le\alpha\le0.96$, whereas significant changes in $\theta_i$ are observed for $\alpha\le0.02$ or $\alpha\ge0.98$.

\begin{figure}
\centering
\includegraphics[width=0.85\linewidth]{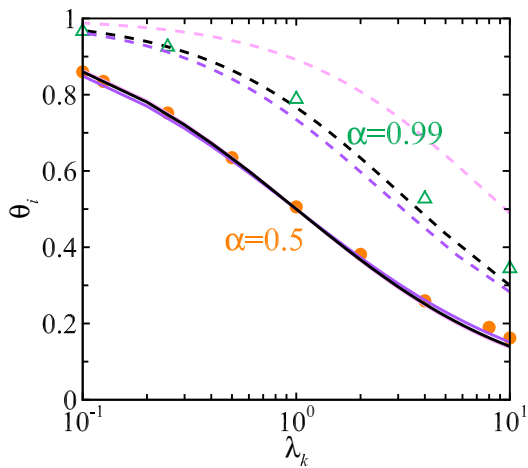}
\caption{\label{ti-k} Temperature at the interface $\theta_i$ as function of $\lambda_k$. The lines represent the full GL theory for the two-layer system (black) in (\ref{gl1})$-$(\ref{gl5}) and the approximate theory in (\ref{scale1}) and (\ref{scale2}) with $\Nun$ vs $\Ra$ scaling exponent of $\gamma=1/4$ (pink) and $\gamma=1/3$ (purple). The symbols and line types are the same as in figure \ref{nu}. 
}
\end{figure}

To explain this $\alpha$ dependence of $\theta_i$, we again use the GL theory to estimate $Nu(Ra,Pr)$ in each of the two layers and then plug the results into eq. (\ref{gl5}) to obtain $\theta_i(\alpha,\lambda_k)$. The results are shown as black lines in figure \ref{ti-a} (for the $\theta_i(\alpha)$ dependence) and figure \ref{ti-k} (for the $\theta_i(\lambda_k)$ dependence), where the GL-predictions for the interfacial temperature $\theta_i(\alpha,\lambda_k)$ very well agree with the numerical results, demonstrating once more the prediction power of the GL theory. 

To obtain a simple relationship for $\theta_i(\alpha,\lambda_k)$, rather than using the full GL theory, an abbreviated approach is to just use an {\it effective} scaling exponent $\gamma$ to approximate the full GL theory, ignoring the crossover between the regimes,
\begin{equation}
    \Nun_j\sim\Ra^\gamma_j,
    \label{scale0}
\end{equation}
where $\gamma$ is constant when $\Ra$ varies over a small range.

Then, by applying eq. (\ref{scale0}) in each layer and imposing the heat flux balance at the interface (\ref{gl3}) and (\ref{gl5}), we get the following approximate scaling law for the regime with both convective layers,
\begin{equation}
    \frac{1}{\theta_i}-1 \sim \lambda_k^\frac{1}{1+\gamma}\,\left(\frac{1}{\alpha}-1\right)^\frac{1-3\gamma}{1+\gamma}.
    \label{scale1}
\end{equation}
We now consider the value of $\gamma$ to vary between $1/4$ and $1/3$, which is reasonable for the classical regime of RB convection \citep{ahl09,loh10,chi12}. Correspondingly, the exponent $(1-3\gamma)/(1+\gamma)$ in (\ref{scale1}) varies from $0.2$ to $0$. This low value reflects the relatively weak $\alpha$ dependence of $\theta_i$ for the regime with two convective layers. 

To validate the simplified relationship (\ref{scale1}), we show the results as pink lines for $\gamma=1/4$ and as purple ones for $\gamma=1/3$ in figure \ref{ti-a}(a) and figure \ref{ti-k}. In both figures, the numerical data and the full GL-predictions are in between the simple relationship of (\ref{scale1}) with $\gamma=1/4$ and $\gamma=1/3$, since a constant effective exponent $\gamma$ is appropriately only within a small range of $\Ra$ while the full GL theory well describes a wide range of $\Ra$.

For the cases with the lower layer being thin enough to be purely conductive (i.e., $\alpha\to1$), we must use $\Nun_1=1$ in the lower layer together with (\ref{scale0}) for the upper layer, which gives
\begin{equation}
    \frac{1-\theta_i}{\theta_i^{(1+\gamma)}} \sim \, \lambda_k \,\alpha^{3\gamma-1}(1-\alpha).
        \label{scale2}
\end{equation}
Although the term $\alpha^{3\gamma-1}$ has weak $\alpha$ dependence (considering $\gamma$ to be between $1/4$ and $1/3$), $\theta_i$ is sensitive to $\alpha$ because of the term $1-\alpha$. Similarly, in the case when the upper layer is purely conductive ($\alpha\to0$), we obtain
\begin{equation}
    \frac{\theta_i}{(1-\theta_i)^{1+\gamma}} \sim \, \lambda_k \,(1-\alpha)^{3\gamma-1}\alpha,
        \label{scale3}
\end{equation}
which also indicates that $\theta_i$ is sensitive to $\alpha$. Furthermore, eqs. (\ref{scale1})$-$(\ref{scale3}) show that $\theta_i$ is sensitive to $\lambda_k$ in all the situations.

We again validate the derived approximate relations (\ref{scale2}) and (\ref{scale3}) using $\gamma=1/4$ and $\gamma=1/3$ in figures \ref{ti-a}(b), \ref{ti-a}(c), and \ref{ti-k}. Good agreements are also achieved in this regime with one conductive layer and one convective layer.

\section{Effects of the deformable interface}
\label{sec4}

\begin{figure}
\centering
\includegraphics[width=\linewidth]{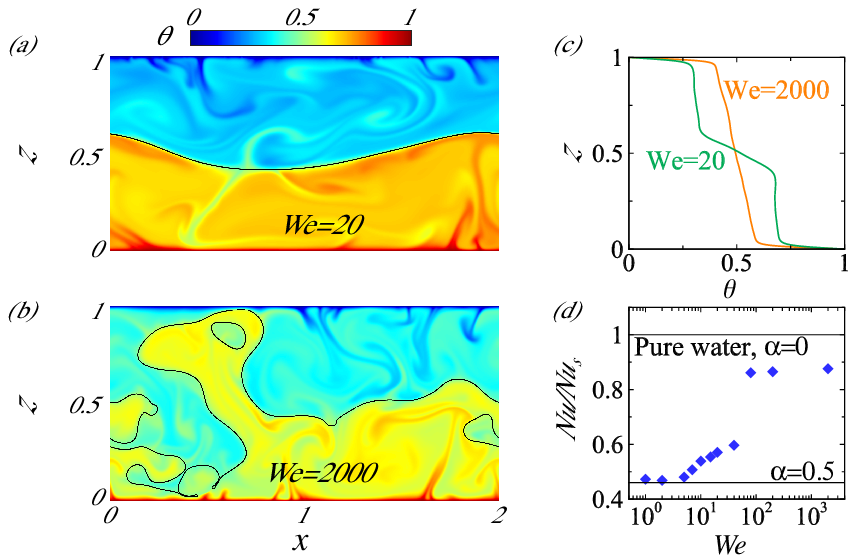}
\caption{\label{we} (a) Snapshots with the wavy interface at $\We=20$ and (b) with the breakup interface at $\We=2000$. (c) Corresponding average temperature profiles. (d) $\Nun/\Nun_s$ as function of $\We$. The lines represent the extended GL theory for the two-layer system, ignoring the interface dynamics. All the cases here are at fixed $\Ra=10^8$, $\alpha=0.5$, and $\lambda_k=1$.}
\end{figure}

In RB convection with two layers, the large-scale circulations are confined within each layer provided that the surface tension is strong enough, i.e., $\We$ small enough. E.g. for $\We=5$ as in our simulations, the interface only deforms slightly as can be seen in figure \ref{3d} and figure \ref{2d}. However, on increasing $\We$, the interface becomes wavy (see figure \ref{we}a) and even breaks up (see figure \ref{we}b). We have derived the criterium for interface breakup in our previous study \citep{liu2021jfm}.

In the wavy case, the large-scale circulation still exists in each layer. In contrast, in the breakup case, the surface tension is too weak and stirring of the fluid layers takes place as evidenced by the vanishment of the sharp temperature jump in the middle of the cell in figure \ref{we}(c).

The Nusselt number also exhibits a highly non-trivial $\We$ dependence, in accordance with the quite different flow dynamics. When the interface remains intact, the values of $\Nun/\Nun_s$ are close to the prediction from the extended GL theory for the two-layer system. When there is a wavy interface, $\Nun/\Nun_s$ becomes slightly larger than the GL prediction. However, when the interface breaks up, $\Nun/\Nun_s$ directly jumps to a saturated value, which is slightly smaller than the value obtained for single-phase flow of the better conducting layer. The reason for the slight difference to the single phase RB case is that although a clear two-layer structure has been washed out, the existence of the heavy droplets of fluid 1 and the light ones of fluid 2 will anyhow slow down the convective flow, hindering the heat flux.

\section{Conclusions and outlook}\label{conc}
Turbulent RB convection with two immiscible fluid layers is numerically investigated for various layer thicknesses ($0.01\le\alpha\le0.99$) and thermal conductivity coefficient ratio ($1\le\lambda_k\le10$) both in 2D and 3D. Two regimes can be identified: In regime 1, two convective layers appear, while in regime 2, one layer is sufficiently thin and dominated by pure conduction because the local $\Ra$ thereof is smaller than the onset value of convection. The heat transfer $\Nun$ is sensitive to the layer thickness $\alpha$ in regime 2 but not in regime 1. To explain the observed $\Nun$ behaviors, we extend the GL theory of single-phase RB convection to the two-layer system. The predictions from the extended GL theory for the overall heat transport well agree with the simulations for various $\lambda_k$ and also with experimental data. Moreover, also the temperature at the interface can be predicted nicely for the two-layer system. We further derived a simple scaling model to explicitly show the relationship between $\theta_i$, $\alpha$, and $\lambda_k$. Finally, the effect of the deformable interface on the heat transfer is studied, where we find that the extended GL theory for the two-layer system still roughly holds, as long as the interface does not break up.

This study showed that the extended GL theory for the two-layer system works for various ratios of the thermal conductivity coefficient in two layers, which is a key property among all material properties, but we also expect that the theory also holds when we vary the ratio of other properties, e.g., the thermal expansion coefficient, the density, the dynamic viscosity, the thermal diffusivity, and the specific heat capacity. The ratios of all properties can be represented by the local dimensionless numbers $\Ra_1$, $\Ra_2$, $\Prn_1$, and $\Prn_2$ in the extended GL theory. It would be also interesting to test this theory for various other global $\Ra$ and $\Prn$, which here were fixed at $\Ra=10^8$ and $\Prn=4.38$.

The results in this work can find many realistic applications, such as in atmospheric oceanography, geophysics, and industry. Our main finding, the layer thickness dependence of heat transport in the two-layer system, also suggests that the two-layer system is not a good choice if one wants to enhance the overall heat transfer. When the two-layer system is unavoidable anyhow, to ensure the maximal heat transfer, the layer with the lower thermal conductivity coefficient must be kept thin enough.

\section*{Acknowledgments}
The work was financially supported by ERC-Advanced Grant under the project no.~$740479$ and the NWO Max Planck Centre ``Complex Fluid Dynamics". We acknowledge PRACE for awarding us access to MareNostrum in Spain at the Barcelona Computing Center (BSC) under the project $2020225335$ and $2020235589$. K. L. Chong acknowledges Shanghai Science and Technology Program under project no. 19JC1412802. This work was also carried out on NWO Domain Science for the use of the national computer facilities.

\section*{Declaration of interests}
The authors report no conflict of interest.



\end{document}